# THERMODYNAMIC MODEL OF THE FLUID SYSTEM $H_2O$-$CO_2$-NaCl-$CaCl_2$ AT *P-T* PARAMETERS OF THE MIDDLE AND LOWER CRUST

Ivanov M. V.

*Institute of Precambrian Geology and Geochronology (IPGG) RAS, St. Petersburg*

*m.v.ivanov@ipgg.ru*

**Abstract.** Based on the earlier obtained equations of state for the ternary systems $H_2O$–$CO_2$–$CaCl_2$ and $H_2O$–$CO_2$–NaCl, an equation of state for the four-component fluid system $H_2O$–$CO_2$–NaCl–$CaCl_2$ is derived in terms of the Gibbs excess free energy. A corresponding numerical thermodynamic model is build. The main part of the numerical parameters of the model coincides with the corresponding parameters of the ternary systems. The NaCl–$CaCl_2$ interaction parameter was obtained from the experimental liquidus of the salt mixture. Similar to the thermodynamic models for $H_2O$–$CO_2$–$CaCl_2$ and $H_2O$–$CO_2$–NaCl, the range of applicability of the model is pressure 1-20 kbar and temperature from 500°C to 1400°C. The model makes it possible to predict the physicochemical properties of the fluid involved in most processes of deep petrogenesis: the phase state of the system (homogeneous or multiphase fluid, presence or absence of solid salts), chemical activities of the components, densities of the fluid phases, and concentrations of the components in the coexisting phases. The model was used for a detailed study of the phase state and activity of water on the $H_2O$–$CO_2$–salt sections when changing the ratio $x_{NaCl}/(x_{NaCl}+x_{CaCl2})$ from 1 to 0. Changes in the composition and density of coexisting fluid phases at a constant activity of water and changes in the total composition of the system are studied. A set of phase diagrams on sections $H_2O$–NaCl–$CaCl_2$ for different mole fractions of $CO_2$ is obtained. Pressure dependencies of the maximal activity of water in the field of coexisting unmixable fluid phases are obtained for several salt compositions of the system. Due to removal of restrictions resulting from a smaller number of components in ternary systems, the thermodynamic behavior of systems with a mixed composition of the salt is significantly differ from the behavior of those with a single salt component.

*Keywords: high temperature, high pressure, fluid, four-component system, phase splitting, $CO_2$, NaCl, $CaCl_2$*

INTRODUCTION

Deep water fluids play an important role in crustal petrogenesis and in the transport of deep matter into the upper crust. Knowledge of the physicochemical properties of deep fluids is an important tool for studying metamorphic, metasomatic and magmatic petrogenesis, the features of the manifestation and development of these global geological processes. Typical components of aqueous fluids are chlorides of alkali or alkaline earth metals and dissolved non-polar gases, in particular $CO_2$. As the deep fluid approaches the surface, the temperature and pressure change significantly, which leads to a change in the physicochemical characteristics of the fluid. As a result of a decrease in pressure, a homogeneous fluid can decompose into immiscible fluid phases with contrasting physical and chemical properties and compositions that differ significantly from the composition of the initial fluid. In addition to the problems of petro- and ore genesis, the knowledge of the quantitative characteristics of deep fluids is necessary for solving a number of geodynamic problems, in particular, the problem of earthquake prediction (Leonov et al., 2006; Kissin, 2009; Rodkin and Rundqvist, 2017; Manning, 2018). Aqueous fluids of the composition $H_2O$–$CO_2$–salt are among the most widespread fluids, participating in the processes of deep metamorphism and metasomatism, transportation of deep-seated ore matter into the upper layers of the Earth's crust (Trommsdorf et al., 1985; Bischoff et al., 1996; Markl, Bucher, 1998; Heinrich et al., 2004; Manning, Aranovich, 2014, 2017). The earliest thermodynamic models of the $H_2O$-$CO_2$-NaCl fluid system for high temperatures and pressures were presented in (Joyce and Holloway, 1993; Duan et al., 1995). Modern numerical thermodynamic models of ternary systems $H_2O$-$CO_2$-salt (Aranovich et al. 2010; Ivanov, Bushmin, 2019, 2021) for PT parameters of the middle and lower crust provide good reproduction of experimental data (Zhang, Frantz, 1989; Kotelnikov, Kotelnikova, 1990; Johnson, 1991; Frantz et al., 1992; Aranovich and Newton, 1996; Shmulovich and Graham, 1999, 2004). All the mentioned thermodynamic models refer to systems containing one particular salt (NaCl or $CaCl_2$). However, aqueous-salt fluids typically contain a mixture of several salts. Fluid inclusions containing a mixture of several salts are a common object of laboratory microthermometry studies (Van den Kerkhof and Hein, 2001). In these studies, in particular, attention is paid to the $H_2O$–NaCl–$CaCl_2$ system (Steelt-MacInnis et al, 2011). Both salts of this system are among the most widespread salts in crustal fluids (Liebscher, 2007). Inclusions, containing $H_2O$–NaCl–$CaCl_2$, may also contain gases, including $CO_2$, or be formed during phase decomposition of the water-gas-salt fluid. Thermodynamic models that describe fluids with a mixed composition of salts under conditions of capture of these inclusions, that is, at sufficiently high temperatures and pressures, are currently absent.

In our previous works (Ivanov, Bushmin, 2019, 2021), numerical thermodynamic models of two ternary fluid systems with water, $CO_2$, and salt components were developed. Both models are valid at temperatures from 500°C to 1400°C and pressures from 1 to 20 kbar. The models were built for NaCl (Ivanov and Bushmin, 2021) and CaCl$_2$ (Ivanov and Bushmin, 2019) salts. In this paper, we present a thermodynamic model of the four-component H$_2$O–CO$_2$–NaCl–CaCl$_2$ system, which is a natural development of our models for the H$_2$O–CO$_2$–NaCl and H$_2$O–CO$_2$–CaCl$_2$ systems.

## THEORETICAL METHOD

### *Gibbs free energy*

The expression presented below for the excess Gibbs free energy of mixing in the four-component system $G^{\text{mix}}$ is based on an approach similar to that used for three-component systems in (Aranovich et al. 2010; Ivanov and Bushmin 2019, 2021). The expression for the entropy part of the Gibbs free energy $G_S$ was obtained by extending the expressions for the corresponding factors in the activities of the components for ternary systems (Aranovich et al. 2010) to the quaternary system. When used for the mole fractions of the components of the notation $x_1 = x_{\text{H}_2\text{O}}$, $x_2 = x_{\text{CO}_2}$, $x_3 = x_{\text{NaCl}}$, $x_4 = x_{\text{CaCl}_2}$, $x_1 + x_2 + x_3 + x_4 = 1$ this term has the form

$$G_S = RT\{(1-x_2)\ln(1-x_2) - [x_1 + (1+\alpha_3)x_3 + (1+\alpha_4)x_4]\ln[x_1 + (1+\alpha_3)x_3 + (1+\alpha_4)x_4] + \\ + x_1\ln x_1 + x_2\ln x_2 + (1+\alpha_3)x_3\ln[(1+\alpha_3)x_3] + (1+\alpha_4)x_4\ln[(1+\alpha_4)x_4]\} \quad (1)$$

The values $\alpha_3$, $\alpha_4$ are the numbers of additional particles formed during the dissociation of the corresponding molecules. $0 \leq \alpha_3 \leq 1$, $0 \leq \alpha_4 \leq 2$. The temperature $T$ is in Kelvin, $R$ is the universal gas constant.

The expression for the contribution to the Gibbs free energy $G_{\text{wg}}$ from the interaction of water and CO$_2$ molecules coincides with the one proposed by (Aranovich et al. 2010; Aranovich 2013) and used in (Ivanov and Bushmin 2019, 2021)

$$G_{\text{wg}} = W_{12}\frac{x_1 x_2 (x_1 + x_2)}{(V_1 x_1 + V_2 x_2)} \quad (2)$$

where $W_{12} = 0.202046$ J·m$^3$/mol, and $V_1$ and $V_2$ are the molar volumes of pure water and carbon dioxide at a given temperature and pressure, respectively.

The Gibbs free energy of the ternary systems H$_2$O–CO$_2$–CaCl$_2$ и H$_2$O–CO$_2$–NaCl (Ivanov and Bushmin, 2019, 2021) contains a term corresponding to the pair interaction of water and salt molecules $W_2 x_1 x_3$. For a quaternary system with two salts, the corresponding expression should include terms corresponding to the interactions H$_2$O–NaCl, H$_2$O–CaCl$_2$, as well as a new term describing the interaction of two salts NaCl–CaCl$_2$

$$G_{\text{wss}} = W_{13}x_1 x_3 + W_{14}x_1 x_4 + W_{34}x_3 x_4 \quad (3)$$

with coefficients $W_{13}$, $W_{14}$, $W_{34}$ for the interactions $H_2O$–NaCl, $H_2O$–$CaCl_2$ и NaCl–$CaCl_2$, respectively. For the $H_2O$-$CO_2$-salt ternary system, the $CO_2$-salt interaction looked like

$$G_{23} = x_2 x_3 [x_2 W_3 + x_3 W_4]/(x_2 + x_3)$$

Similarly, the corresponding terms for the quaternary system have the form

$$G_{gs} = x_2 x_3 [x_2 W_{23} + x_3 W_{32}]/(x_2 + x_3) + x_2 x_4 [x_2 W_{24} + x_4 W_{42}]/(x_2 + x_4) \qquad (4)$$

The Gibbs free energy of mixing for ternary systems $H_2O$-$CO_2$-salt also contained a term describing the interaction of three types of molecules in the system $W_5 x_1 x_2 x_3$, i.e. water-$CO_2$-salt. For the quaternary system, the corresponding expression consists of two terms – for interactions of $H_2O$-$CO_2$-NaCl and $H_2O$-$CO_2$-$CaCl_2$

$$G_{wgs} = W_{123} x_1 x_2 x_3 + W_{124} x_1 x_2 x_4 \qquad (5)$$

Thus, the complete expression for the Gibbs free energy of mixing $G^{mix}$ of the four-component fluid under consideration has the form

$$G^{mix}(x_1, x_2, x_3, x_4) = G_S + G_{wg} + G_{wss} + G_{gs} + G_{wgs} \qquad (6)$$

At $x_3 = 0$ or $x_4 = 0$ equation (1) transforms into equations for the ternary systems $H_2O$-$CO_2$-$CaCl_2$ (Ivanov, Bushmin, 2019) or $H_2O$-$CO_2$-NaCl (Ivanov, Bushmin, 2021). Knowing the Gibbs free energy (6) allows us to determine other thermodynamic functions of the quaternary system $H_2O$-$CO_2$-NaCl-$CaCl_2$. The most important of them are chemical potentials, component activities, and fluid density. Formulas for these quantities are given in the Appendix.

Expression (6) describes the excess free energy of mixing for a homogeneous four-component fluid. In a more general case, the system we are considering can contain up to two fluid phases, each of which is described by equation (6), as well as solid phases of salts. The excess Gibbs free energy of such a generally multiphase system has the form

$$G^M(x_1, x_2, x_3, x_4) = f_1 G^{mix}(y_{11}, y_{21}, y_{31}, y_{41}) + f_2 G^{mix}(y_{12}, y_{22}, y_{32}, y_{42}) + \Delta\mu_{s3} x_{s3} + \Delta\mu_{s4} x_{s4} \qquad (7)$$

where $\Delta\mu_{s3}$, $\Delta\mu_{s4}$ are the values of the change in the chemical potentials of NaCl and $CaCl_2$ during the solid-liquid transition, $x_{s3}$, $x_{s4}$ are the mole fractions of solid salts in the multiphase system, $f_1$ and $f_2$ are the mole fractions of two, in the general case, different fluid phases,

$$y_{ij} = x_{ij} / f_j$$
$$f_j = x_{1j} + x_{2j} + x_{3j} + x_{4j}$$
$$x_{12} = x_1 - x_{11}$$
$$x_{22} = x_2 - x_{21}$$
$$x_{32} = x_3 - x_{31} - x_{s3}$$
$$x_{42} = x_4 - x_{41} - x_{s4}$$

The phase state of the system for given mole fractions of the components $(x_1, x_2, x_3, x_4)$ is determined by numerical minimization of the quantity $G^M$ with respect to the variables $(x_{11}, x_{21}, x_{31}, x_{41}, x_{s3}, x_{s4})$, provided that all the quantities $x_{ij}$ are non-negative.

## NUMERICAL PARAMETERS OF THE MODEL

The parameters $W_{ij}$ in equation (1) are assumed to depend on the temperature $T$ and pressure $P$, but not on the mole fractions of the components. The $P$-$T$ form of the dependences of the parameters $W_{ij}$, $W_{ijk}$ and the corresponding numerical values are given in (Ivanov, Bushmin, 2019, 2021). These parameters, obtained by fitting the experimental data, are retained in the thermodynamic model of the four-component system presented in this paper. The excess free energy of the multiphase system (7) also contains the functions $\Delta \mu_{s3}(P,T), \Delta \mu_{s4}(P,T)$. The form of these functions and the corresponding numerical parameters are also given in (Ivanov, Bushmin, 2019, 2021).

The only new parameter of the model of a four-component system, which was not obtained in previous works, is the parameter $W_{34}$, which describes the effect of the interaction of NaCl and CaCl2 particles in the fluid. In the case $x_1 = x_2 = 0$, equation (6) retains only the entropy terms (1) and the term $W_{34}x_3x_4$. In this case, equation (7) turns out to be an equation of state for a mixture of NaCl and CaCl$_2$ salts, which are either in a molten or in a solid state. For $P = 1$ bar and temperatures above 500°C, our equations give $\alpha_3 = \alpha_4 = 0$, which provides the usual form of the free energy entropy term for a mixture of two substances NaCl and CaCl$_2$. Thus, equation (7) should describe the solid-liquid transition in a mixture of NaCl and CaCl$_2$ salts. Experimental data on liquidus in the NaCl-CaCl$_2$ system at $P = 1$ bar are well known (Seltveit and Flood, 1958; Chartrand and Pelton, 2001). The liquidus has a simple form, typical of mixtures that do not form intermediate compounds. The tit of values of the parameter $W_{34}$ makes it possible to reproduce the experimental liquidus of NaCl and CaCl$_2$. In Fig. 1 we provide a comparison of the liquidus obtained with our equation of state at $W_{34} = -10.3$ kJ/mol with experimental results (Seltveit, Flood, 1958). At $W_{34} = -10.3$ kJ/mol, the eutectic point corresponds to a temperature of 502°C and $x_{\text{NaCl}} = 0.482$, which is in very good agreement with numerous experimental data (Chartrand and Pelton, 2001). Parameter $W_{34}$ is independent of pressure. However, the pressure dependences of the values $\Delta\mu_{s3}$, $\Delta\mu_{s4}$ make it possible to obtain the theoretical liquidus in the NaCl-CaCl$_2$ system at higher pressures. An example of such a liquidus for $P = 5$ kbar is given in Fig. 1. The obtained value $W_{34} = -10.3$ kJ/mol, in combination with the parameters from (Ivanov, Bushmin, 2019, 2021), forms a complete set of parameters for our thermodynamic model of the four-component system H$_2$O-CO$_2$-NaCl-CaCl$_2$.

## RESULTS

*Phase diagrams. Their evolution with a change in the composition of the salt component*

The phase diagrams of ternary systems are comprehensively represented as composition triangles with pure components at the vertices of the triangles. Phase fields are two-dimensional regions inside a triangle, and the boundaries of these regions are lines, in particular the solvus line. A similar phase diagram for a quaternary system should be a tetrahedron with volume phase fields and two-dimensional surfaces – the boundaries of these fields. A meaningful and more or less accurate presentation of the results in this form is difficult. It is more convenient to represent phase diagrams on plane sections of the tetrahedron of compositions, in particular, on sections with a fixed composition of the salt part of the system $r_{NaCl} = x_{NaCl}/x_{salt}$, where $x_{salt} = x_{NaCl} + x_{CaCl_2}$. The phase diagrams on these sections turn out to be triangles in coordinates $x_{H_2O} - x_{CO_2} - x_{salt}$ and have a form similar to the phase diagrams of ternary systems. At $r_{NaCl} = 0$ or $r_{NaCl} = 1$ these phase diagrams are phase diagrams of the H$_2$O–CO$_2$–CaCl$_2$ or H$_2$O–CO$_2$–NaCl ternary systems, respectively.

In Fig. 2, we show a series of phase diagrams $x_{H_2O} - x_{CO_2} - x_{salt}$ corresponding to a smooth change in the composition of the salt component of the system from pure NaCl (Fig. 2a) to pure CaCl$_2$ (Fig. 2l). All the diagrams in Fig. 2 are plotted for typical *P-T* conditions of the amphibolite facies metamorphism $T = 600$°C and $P = 5$ kbar. The edge ternary systems in Figs. 2a and 2l show a complex phase pattern consisting of five different fields: 1. Homogeneous fluid; 2. Two coexisting fluid phases; 3. Two fluid phases coexisting with solid salt; 4. Brine coexisting with solid salt; 5. CO$_2$-rich fluid coexisting with solid salt. In the extensive *field-3*, there are three coexisting phases, which ensures the constancy of the activities of the components throughout the field ($a_{H_2O} = 0.367$, $a_{CO_2} = 0.783$, $a_{NaCl} = 0.345$ at $r_{NaCl} = 1$ and $a_{H_2O} = 0.290$, $a_{CO_2} = 0.837$, $a_{CaCl_2} = 0.531$ at $r_{NaCl} = 0$).

The phase diagram Fig. 2g for a fluid with equal mole fractions of NaCl and CaCl$_2$ contains only two-phase fields, namely, the field of a homogeneous fluid and the field of two coexisting fluid phases. Within some quantitative changes, the phase diagrams for $0.24 \leq r_{NaCl} \leq 0.67$ look similar. The disappearance of fields containing solid salt is due to the greater solubility of salt of a mixed composition compared to its pure components. However, the differences between fluids with a single pure salt component and mixed salts are deeper.

Comparison of Figs. 2a and 2b, as well as Figs. 2l and 2k shows that the addition of even 0.1% $CaCl_2$ to pure NaCl in the first case and 0.2% NaCl to pure $CaCl_2$ in the second case fundamentally changes the phase diagram. *Field-3* with three coexisting phases increases drastically, while *field-5* of the $CO_2$-rich fluid coexisting with solid salt is reduced to negligible size. Field-3 in Fig. 2b and Fig. 2k is no longer a region of constant water activity as in the case of pure salt. In contrast to the case of pure salts, this field is intersected by $a_{H_2O}$ isolines. For a system with pure NaCl, the water activity in the region of coexistence of two fluid phases (*fields-2* and *3*) could not be less than $a_{H_2O} = 0.367$. The addition of even fractions of a percent of the second salt makes it possible for the two fluid phases to coexist at an arbitrarily low water activity. The same thing happens when small amounts of NaCl are added to pure $CaCl_2$. The transition to an even more mixed composition of the salt (Figs. 2c-2f and Figs. 2k-2h) leads to a rearrangement of the position of water activity isolines, a gradual reduction in *field-3* and a change in the position of the solvus.

*Compositions and densities of coexisting fluid phases*

A significant difference between a quaternary fluid system and ternary systems is associated with the composition of coexisting fluid phases. In the ternary water-gas-salt system, the activity isolines of the components are curves in the field of a homogeneous fluid. If they cross the solvus and pass into the field of two coexisting fluids, they become straight lines – tie lines connecting the two points of intersection with the solvus. The compositions of the two coexisting fluid phases along the tie lines are constant. One of these phases has a composition of the homogeneous fluid at the upper intersection point. The composition of the other phase coincides with the composition of the homogeneous fluid at the lower intersection point. In a four-component system, the difference between the interaction energies of NaCl and $CaCl_2$ with water and $CO_2$ leads to a redistribution of salts between coexisting fluid phases with a change in the total composition of the system. As a result, the compositions of the coexisting fluid phases are not constant along the isolines of constant activities. These isolines are not tie lines and, in general, they are not straight lines. An example of redistribution of salt components along the line of constant water activity is given in Fig. 3. This figure presents data characterizing the compositions and density of two fluid phases, denoted as *f1* and *f2*, along the isoline $a_{H_2O} = 0.6$ of Fig. 2f (the total composition of the system corresponds to $r_{NaCl} = 0.8$) depending on the total mole fraction of $CO_2$ in the system ($x_{CO2, total}$ in the figure). In the Fig. 3a, ratios $r_{NaCl}$ for phases *f1* and *f2* are shown. For the homogeneous fluid, $r_{NaCl} = 0.8$. After the isoline crosses the solvus at $x_{CO_2} = 0.122$, the second fluid phase *f2* appears (green curve). This phase is in equilibrium with the initial phase at $a_{H_2O} = 0.6$, but the ratio of the mole fractions of salts in it is very different from the general one $r_{NaCl} = 0.8$. This second phase is enriched in NaCl due to the negligible loss of NaCl in the original fluid (blue line). In the course of following the isoline, the two fluid phases exchange salts and other components. Further, at $x_{CO_2} = 0.474$ the initial fluid phase *f1* disappears, while the phase *f2* comes to the ratio of mole fractions of salts of a homogeneous fluid $r_{NaCl} = 0.8$. Fig. 3b shows the total salinities of the same fluid phases. The mole fractions of $CO_2$ in the coexisting phases are given in Fig. 3c. The densities of the phases are given in Fig. 3d. The *f1* phase is predominantly water-salt with a small amount of carbon dioxide, while *f2* is a water-carbon dioxide phase with a very low salt content. In the case of a ternary system with one salt, the coexisting phases would also be water-salt and water-carbon dioxide, but their compositions and densities would be constant on the line of constant water activity (tie line).

*Phase diagrams in $H_2O$–NaCl–$CaCl_2$ coordinates*

In addition to the phase diagrams of Fig. 2 on sections with a constant ratio of mole fractions of two salts, in Fig. 4 we present the phase diagrams of the water-salt system at fixed mole fractions of carbon dioxide. This is the phase diagram of the three-component system water-two salts in Fig. 4a and phase diagrams in Figs. 3b, 3c, and 3d for nonzero $x_{CO_2}$. In the absence of carbon dioxide in the system (Fig. 4a), phase decomposition of the fluid does not occur. Therefore, a homogeneous fluid containing $H_2O$, NaCl, and $CaCl_2$ is present for all mole fractions of these components that do not vanish. At a high concentration of one of the salts and a relatively low mole fraction of water, a homogeneous fluid coexists with the solid phase of the corresponding salt.

The addition of even a small amount of carbon dioxide to the system (Fig. 4b) leads to the appearance of a wide field of the fluid, which decomposes into water-carbon dioxide and water-salt phases. A further increase in the proportion of CO2 in the system (Fig. 4c) leads to a serious reduction in the field of a homogeneous fluid and to its almost complete disappearance at $x_{CO_2} = 0.4$ (Fig. 4d).

*Restrictions on the value of water activity in the area of fluid splitting into two phases*

As can be seen from the phase diagrams presented in Figs. 2, in the four-component system we are considering, the chemical activity of water in the region of a two-phase fluid has no lower limit and can drop to values close to zero. In this point, the thermodynamic behavior of a quaternary system with a mixed composition of the salt component differs from the behavior of similar ternary systems with a single salt component. However, in the system $H_2O–CO_2–NaCl–CaCl_2$, as well as in ternary systems, the chemical activity of water in the region of fluid splitting into two coexisting fluid phases is limited from above. The value of this limiting activity depends on the temperature, pressure, and the ratio of the mole fractions of $CO_2$ and $CaCl_2$ in the composition of the salt. In Fig. 5a, we show solvuses for the $H_2O–CO_2–NaCl–CaCl_2$ fluid with equal mole fractions of NaCl and $CaCl_2$ for several pressures from 1 kbar to 20 kbar at temperature 700°C. As the pressure increases, the field of a homogeneous fluid expands, and the solvuses, together with the critical points at which the water activity takes on values, that are maximum possible for a fluid in a heterogeneous field, shift to the region of higher salt and carbon dioxide concentrations. An increase in the molar fractions of the salts and $CO_2$ leads to a decrease in water activity at critical points. Figure 5b shows the pressure dependences of this limiting value of $a_{H_2O}$ at $T = 700°C$ for the two extreme systems with pure salts and for the intermediate system with $r_{NaCl} = 0.5$. In all three systems, an increase in pressure leads to an expansion of the homogeneous fluid field and, accordingly, a decrease in water activity at the critical point of the two-phase fluid field. With an increase in pressure from 4 kbar and above, the drop in the activity of water with salt in the form of pure $CaCl_2$ is the sharpest. With an increase in the proportion of the NaCl salt component in the composition, the dependence $a_{H_2O}(P)$ becomes less pronounced.

As can be seen from Fig. 5a, an increase in pressure leads to an increase in the mole fraction of salt in the composition of the fluid at the critical point. At the same time, there is a significant increase in the molar fraction of carbon dioxide. As a result, the minimum possible ratio $x_{salt}/(x_{salt} + x_{H_2O})$ in the water-salt part of the two-phase fluid increases to a greater extent than the mole fraction of salt in the total composition of the fluid. The widely used method of microthermometry (Van den Kerkhof, Hein, 2001) used to study natural inclusions makes it possible to determine exactly the value of $x_{salt}/(x_{salt} + x_{H_2O})$. The value of this ratio at the critical point as a function of pressure is shown in Fig. 5c for three ratios $x_{NaCl}/x_{salt}$.

## CONCLUSIONS

In this work, the following results are obtained: **1.** An equation of state of the $H_2O–CO_2–NaCl–CaCl_2$ fluid system, expressed in terms of the excess Gibbs free energy, is derived. The equation of state is valid at temperatures from 500°C to 1400°C and pressures from 1 kbar to 20 kbar. This equation of state is a generalization of the equations of state of the ternary systems $H_2O–CO_2–CaCl_2$ and $H_2O–CO_2–NaCl$ obtained in (Ivanov, Bushmin, 2019, 2021). In the absence of one of the salts in the system, the equation of this work turns into the equations of state of the works (Ivanov, Bushmin, 2019, 2021); **2.** Formulas for the chemical activities of the components have been obtained, corresponding to the derived equation for the excess Gibbs free energy; **3.** Most of the numerical parameters of the equation were obtained in previous studies of the $H_2O$–$CO_2$–NaCl and $H_2O$–$CO_2$–$CaCl_2$ edge systems. The parameter for NaCl–$CaCl_2$ interaction was obtained from experimental data on the liquidus of a mixture of these salts; **4.** Our equation of state makes it possible to calculate most of the thermodynamic characteristics for arbitrary mole fractions of the components; **5.** This paper presents a series of phase diagrams of the system at a fixed ratio of the mole fractions of NaCl and $CaCl_2$. Phase diagrams of this kind, which look similar to phase diagrams for ternary systems, make it possible to trace changes in the thermodynamics of the $H_2O$–$CO_2$–salt system with a change in the composition of the salt. Along with features that depend on the properties of the NaCl–$CaCl_2$ pair and are specific to our model, such a study makes it possible to reveal fundamental differences between ternary and four-component systems; **6.** For given *P* and *T*, both in ternary and quaternary systems there are upper limits of water activity in the field of coexistence of two fluid phases. In case *P-T* conditions allow the existence of a solid salt phase, there is also a lower limit for water activity in the field of coexistence of two fluid phases. This effect is absent in four-component fluid systems with two salts. The activity of water in the field of a two-phase fluid in a four-component system with a mixed salt composition can drop to zero. **7.** In the field of a four-component fluid, split into two coexisting fluid phases, the lines of constant activities of the components are not tie lines, that is, lines on which the compositions of coexisting phases are constant. When moving along these lines, the components are redistributed between coexisting phases. An example of a change in the composition and density of coexisting fluid phases with a change in the total composition of the system and a fixed constant water activity was studied in detail; **8.** A series of phase diagrams of the system was obtained in the coordinates $H_2O$–NaCl–$CaCl_2$ at fixed mole fractions of $CO_2$. The diagrams contain the phase fields of two solid salts coexisting with both homogeneous and two-phase fluids. The changes in the fields of a homogeneous and two-phase fluid with a change in the mole fraction of $CO_2$ were traced; **9.** Dependences on pressure for the maximum activity of water in the region of coexistence of immiscible fluid phases were obtained for different salt composition of the system.

Comparison of the obtained results with similar results for fluids with one salt (Aranovich et al., 2010; Ivanov, Bushmin, 2019, 2021) shows that the thermodynamic behavior of a quaternary system with a mixed composition of the salt component differs significantly from the behavior of edge ternary systems with one salt component due to the removal of restrictions arising from a smaller number of components in ternary systems. This qualitative conclusion is not specific to the thermodynamic model constructed by us and should presumably remain valid for other possible models of the considered and similar systems. The use of numerical thermodynamic models of multicomponent fluid systems expands the possibilities of theoretical analysis of geological processes involving fluids, and provides a new tool for interpreting the results of studies of fluid inclusions.

The computer program that performs calculations according to the above thermodynamic model is available at https://www.dropbox.com/sh/70xaght7deludws/AAA5QygWCrr4sGFqxQpx1b24a?dl=0 .

*Financial sources.* The work was carried out in the framework of the research theme of IPGG RAS FMUW-2021-0002.

# APPENDIX

The chemical potentials of the fluid components corresponding to the Gibbs free energy of mixing (1)-(6) have the form

$$\mu_1 = RT\{\ln(1-x_2) + \ln x_1 - \ln(x_1 + (1+\alpha_3)x_3 + (1+\alpha_4)x_4)\}$$
$$+ W_{12}\frac{x_2(V_1x_1^2 + 2V_2x_1x_2 + V_2x_2^2 - V_1x_1^3 - V_2x_1^2x_2 - V_2x_1x_2^2 - V_1x_1^2x_2)}{(V_1x_1 + V_2x_2)^2}$$
$$+ W_{13}(1-x_1)x_3 + W_{14}(1-x_1)x_4 - W_{34}x_3x_4$$
$$- x_2x_3(W_{23}x_2 + W_{32}x_3)/(x_2+x_3) - x_2x_4(W_{24}x_2 + W_{42}x_4)/(x_2+x_4)$$
$$+ W_{123}x_2x_3(1-2x_1) + W_{124}x_2x_4(1-2x_1)$$

$$\mu_2 = RT \ln x_2$$
$$+ W_{12}\frac{x_1(V_2x_2^2 + 2V_1x_1x_2 + V_1x_1^2 - V_2x_2^3 - V_1x_1x_2^2 - V_1x_1^2x_2 - V_2x_1x_2^2)}{(V_1x_1 + V_2x_2)^2}$$
$$- W_{13}x_1x_3 - W_{14}x_1x_4 - W_{34}x_3x_4$$
$$+ W_{23}x_2x_3(x_2 - x_2^2 - x_2x_3 + 2x_3)/(x_2+x_3)^2 + W_{32}x_3^2(-x_2^2 + x_3 - x_2x_3)/(x_2+x_3)^2$$
$$+ W_{24}x_2x_4(x_2 - x_2^2 - x_2x_4 + 2x_4)/(x_2+x_4)^2 + W_{42}x_4^2(-x_2^2 + x_4 - x_2x_4)/(x_2+x_4)^2$$
$$+ W_{123}x_1x_3(1-2x_2) + W_{124}x_1x_4(1-2x_2)$$

$$\mu_3 = RT\{(1+\alpha_3)\ln[(1+\alpha_3)x_3] + \ln(x_1 + x_3 + x_4) - (1+\alpha_3)\ln[x_1 + (1+\alpha_3)x_3 + (1+\alpha_4)x_4]\}$$
$$- W_{12}\frac{x_1x_2(x_1+x_2)}{(V_1x_1 + V_2x_2)}$$
$$+ W_{13}(1-x_3)x_1 + W_{34}(1-x_3)x_4 - W_{14}x_1x_4$$
$$+ W_{23}x_2^2(x_2 - x_2x_3 - x_3^2)/(x_2+x_3)^2 + W_{32}x_2x_3(2x_2 + x_3 - x_3^2 - x_2x_3)/(x_2+x_3)^2 -$$
$$- W_{24}x_2^2x_4/(x_2+x_4) - W_{42}x_2x_4^2/(x_2+x_4)$$
$$+ W_{123}x_1x_2(1-2x_3) - 2W_{124}x_1x_2x_4$$

$$\mu_4 = RT\{(1+\alpha_4)\ln[(1+\alpha_4)x_4] + \ln(x_1 + x_3 + x_4) - (1+\alpha_4)\ln[x_1 + (1+\alpha_3)x_3 + (1+\alpha_4)x_4]\}$$
$$- W_{12}\frac{x_1x_2(x_1+x_2)}{(V_1x_1 + V_2x_2)}$$
$$+ W_{14}(1-x_4)x_1 + W_{34}(1-x_4)x_3 - W_{13}x_1x_3$$
$$+ W_{24}x_2^2(x_2 - x_2x_4 - x_4^2)/(x_2+x_4)^2 + W_{42}x_2x_4(2x_2 + x_4 - x_4^2 - x_2x_4)/(x_2+x_4)^2 -$$
$$- W_{23}x_2^2x_3/(x_2+x_3) - W_{32}x_2x_3^2/(x_2+x_3)$$
$$+ W_{124}x_1x_2(1-2x_4) - 2W_{123}x_1x_2x_3$$

Chemical potentials for pure components $\mu_{i0}$, i.e. for $x_i = 1$ are equal to zero

$$\mu_{10} = \mu_{20} = \mu_{30} = \mu_{40} = 0$$

Component activities are calculated using common formulas

$$a_i = \exp\left(\frac{\mu_i - \mu_{i0}}{RT}\right)$$

The density of a fluid can be obtained from its molar volume. The latter is calculated as

$$V = \left(\frac{\partial G}{\partial P}\right)_{T,\{x_i\}}$$

Van den Kerkhof, A.M., Hein, U.F., Fliud inclusion petrography, *Lithos*, 2001, vol. 55, pp. 27–47.

Zhang, Y.-G., Frantz, J.D., Experimental determination of the compositional limits of immiscibility in the system $CaCl_2$-$H_2O$-$CO_2$ at high temperatures and pressures using synthetic fluid inclusions, *Chem. Geol.*, 1989, vol. 74, pp. 289–308.

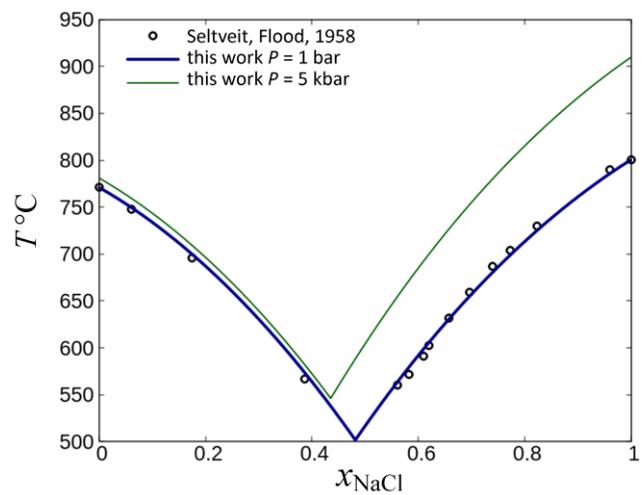

**Fig. 1.** Liquidus in the NaCl-CaCl$_2$ system. Circles are experimental results (Seltveit and Flood, 1958) for $P = 1$ bar. The curves are obtained in our model for $P = 1$ bar and $P = 5$ kbar.

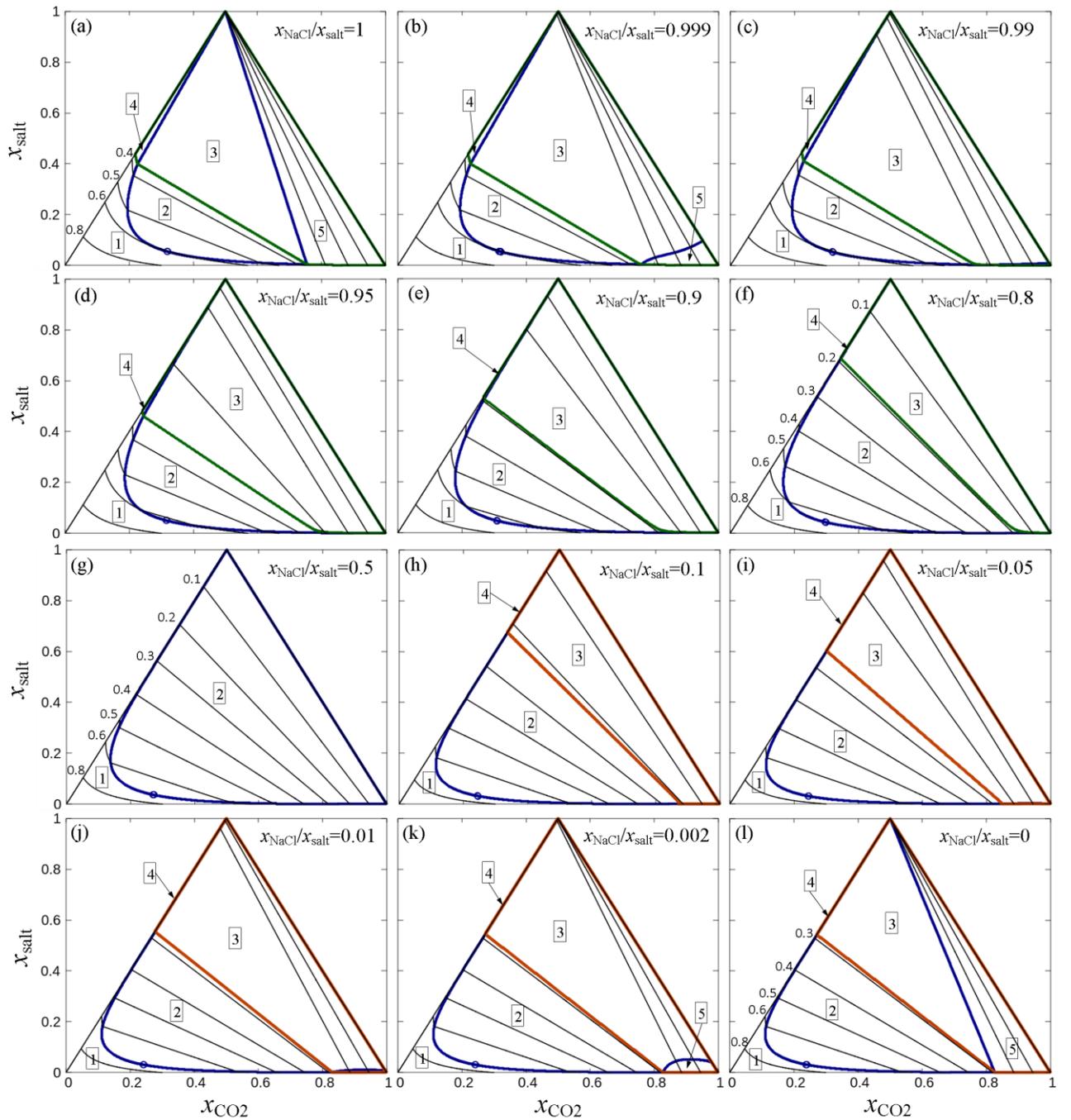

**Fig. 2.** Evolution of the phase diagrams of the H$_2$O-CO$_2$-salt system with a change in the ratio $x_{NaCl}/x_{salt}$, $x_{salt}$. $T$=600°C, $P$=5 кбар. The bold blue line is the boundary of the field of two coexisting fluid phases (solvus). Open circles indicate critical points of the two-phase fluid field. Bold green and orange lines indicate the boundaries of the fields of existence of NaCl (a)–(f) and CaCl2 (h)–(l) solid phases, respectively. Numbers in frames denote areas (fields) of different phase composition: 1 – homogeneous fluid; 2 – two coexisting fluid phases; 3 – two fluid phases coexisting with solid salt ((a)-(f) – NaCl and (h)-(l) – CaCl$_2$); 4 - brine coexisting with solid salt; 5 – CO2 rich fluid coexisting with solid salt. Thin black lines are isolines of chemical activity of water. The activity values are $a_{H_2O} = 0.1, 0.2, 0.3, 0.4, 0.5, 0.6, 0.8$ in the sequence of lines from top to bottom and from right to left (partially indicated by numbers in Fig. 2a, 2f, 2g, 2l).

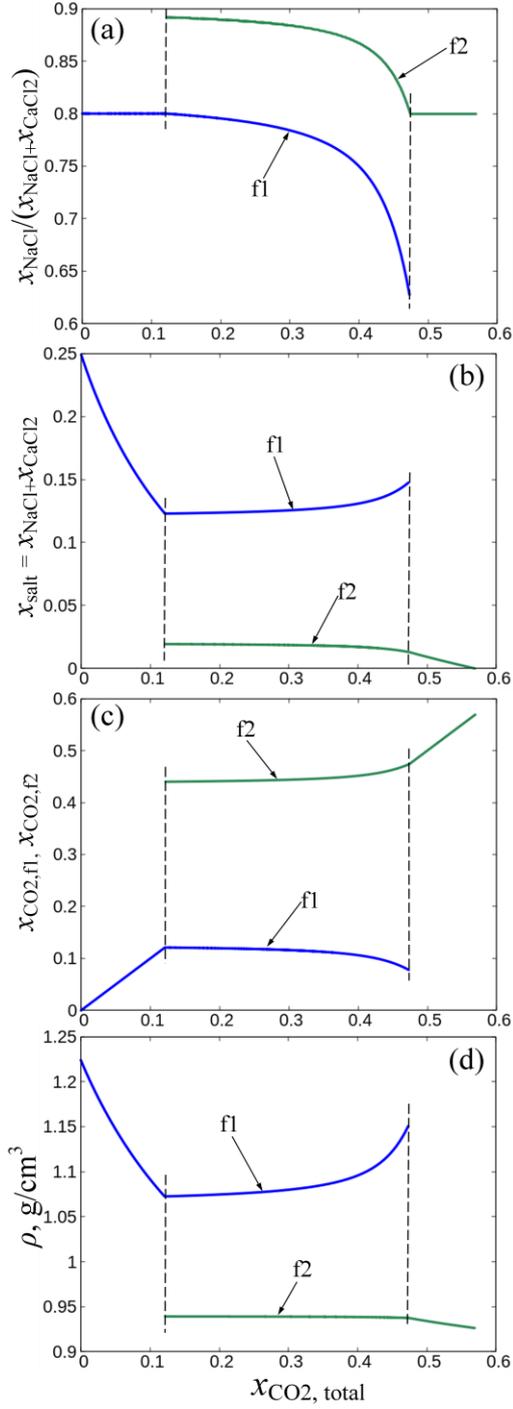

**Fig. 3.** (a) The $x_{NaCl}/(x_{NaCl}+x_{CaCl2})$ ratio in coexisting fluid phases along the isoline $a_{H_2O} = 0.6$ at $T = 600°C$, $P = 5$ kbar and total ratio $x_{NaCl}/(x_{NaCl} + x_{CaCl_2}) = 0.8$ (Fig. 2e). The boundaries of the region of coexistence of two fluid phases are indicated by vertical dashed lines. (b) Total salinities of the fluid phases on the isoline $a_{H_2O} = 0.6$. (c) Mole fraction of $CO_2$ in coexisting fluid phases. (d) Fluid phase densities.

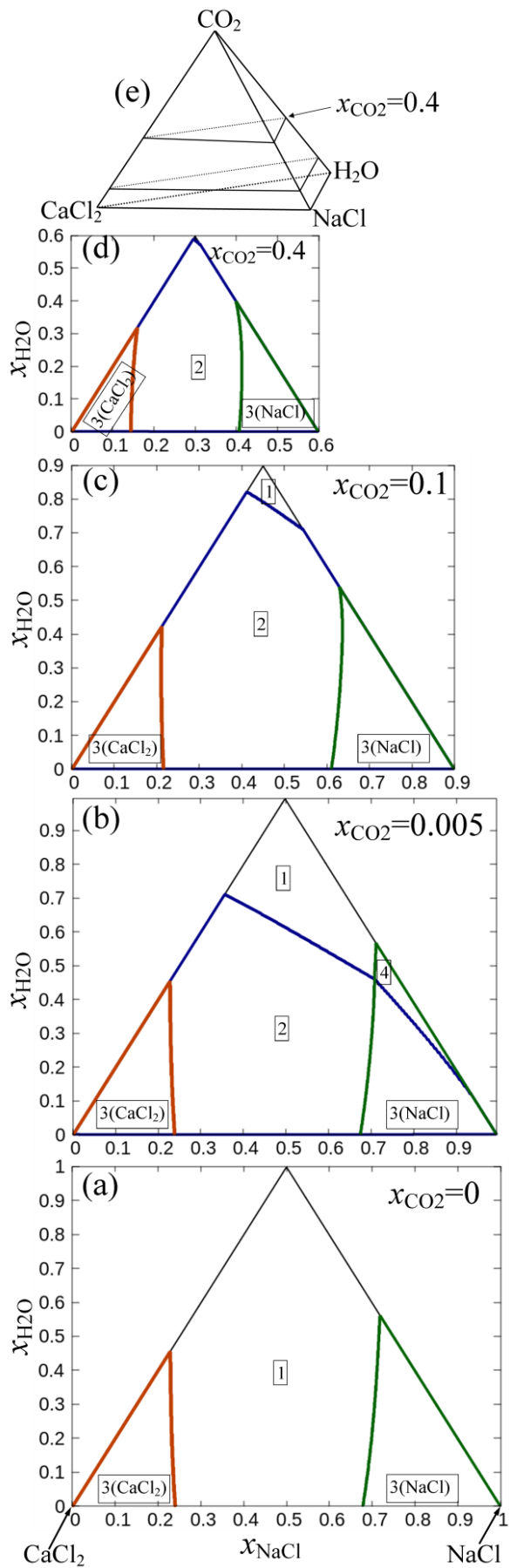

**Fig. 4.** Phase diagrams on sections with a constant CO$_2$ mole fraction, $T = 600°C$, $P = 5$ kbar. For designations of phase fields, see Fig. 2. In *field-3* designations (two fluid phases coexisting with a solid salt), the composition of the solid salt (NaCl or CaCl$_2$) is additionally indicated. (a) $x_{CO_2} = 0$; (b) $x_{CO_2} = 0.005$; (c) $x_{CO_2} = 0.1$; (d) $x_{CO_2} = 0.4$; (e) layout of sections in the tetrahedron of compositions.

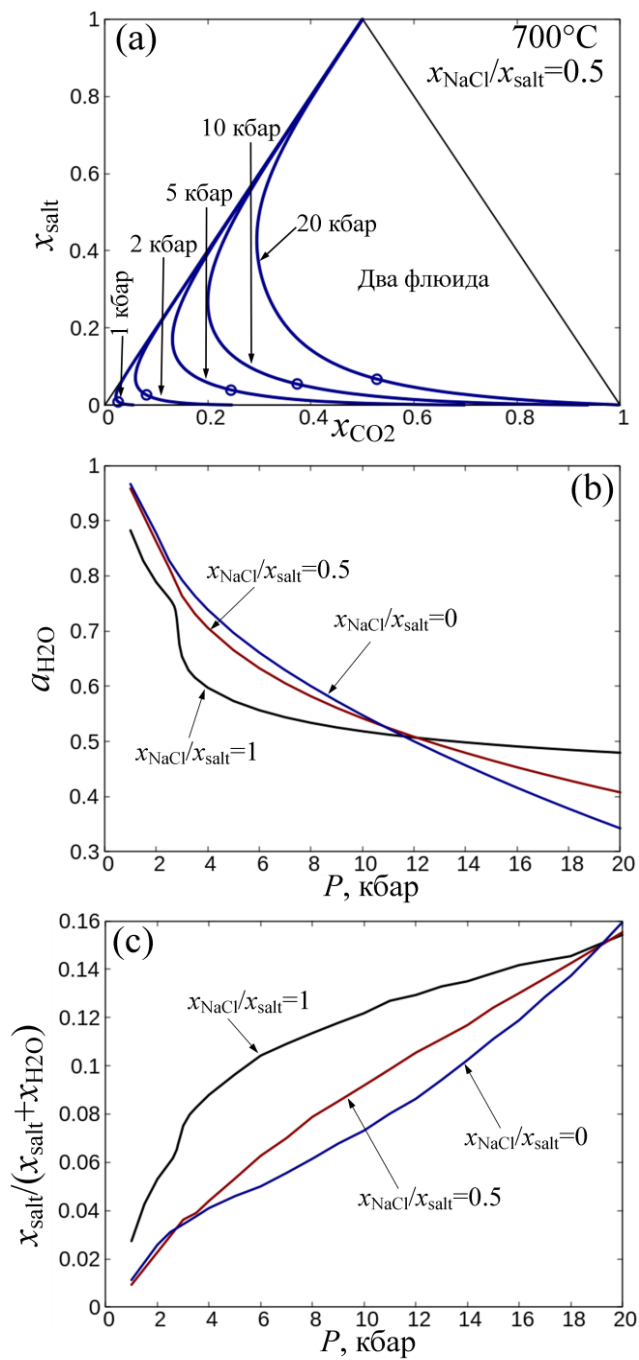

**Fig. 5.** (a) Solvuses (blue curves) and critical points (open circles) for the $H_2O$-$CO_2$-NaCl-$CaCl_2$, $x_{NaCl}/x_{salt} = 0.5$ fluid at $T = 700°C$ and pressures from 1 kbar to 20 kbar. (b) Maximum water activity in the region of coexistence of two fluid phases at three different salt compositions of the fluid system as a function of pressure at $T = 700°C$. (c) The minimum salinity of the water-salt component of the fluid, at which its decomposition into two phases is possible.